\documentclass[journal]{IEEEtran}

\usepackage{amsmath,amsfonts}
\usepackage{algorithm,algpseudocode}
\usepackage{array}
\usepackage[caption=false,font=normalsize,labelfont=sf,textfont=sf]{subfig}
\usepackage{textcomp}
\usepackage{stfloats}
\usepackage{url}
\usepackage{verbatim}
\usepackage{graphicx}
\usepackage{tabularx}
\usepackage{multirow}
\usepackage{booktabs}
\usepackage{xurl}
\usepackage{mathrsfs}
\usepackage[table]{xcolor}
\usepackage{color}

\usepackage{flushend}

\usepackage{balance}

\usepackage{amssymb}

\newcolumntype{Y}{>{\centering\arraybackslash}X}

\definecolor{softBlue}{HTML}{C2E0FF}
\definecolor{softSoftBlue}{HTML}{E6F2FF}

\begin{document}
\title{\mbox{SELEBI}: Percussion-aware Time Stretching via \\Selective Magnitude Spectrogram Compression \\by Nonstationary Gabor Transform}
\author{Natsuki~Akaishi,~\IEEEmembership{Student~Member,~IEEE}, Nicki~Holighaus,~\IEEEmembership{Member,~IEEE}, Kohei~Yatabe,~\IEEEmembership{Member,~IEEE}
\thanks{Manuscript received XXXX XX, XXXX; revised XXXXX XX, XXXX; accepted XXXXX XX, XXXX. Date of publication XXXXX XX, XXXX; date of current version XXXXX XX, XXXX. The associate editor was XXXXX XXXX. \\
Natsuki Akaishi and Kohei Yatabe are with Tokyo University of Agriculture and Technology,
Tokyo 184-8588, Japan (e-mail: natsu61aka@gmail.com;  yatabe@go.tuat.ac.jp). \\
Nicki Holighaus is with Acoustics Research Institute, Austrian Academy of Sciences, Wohllebengasse 12–14, 1040 Vienna, Austria (nicki.holighaus@oeaw.ac.at)}}

\markboth{IEEE/ACM TRANSACTIONS ON AUDIO, SPEECH, AND LANGUAGE PROCESSING,~Vol.~XX, No.~X, XXXX~20XX}
{AKAISHI \MakeLowercase{\textit{et. al.}}: TITLE}

\maketitle

\begin{abstract}

Phase vocoder-based time-stretching is a widely used technique for the time-scale modification of audio signals. However, conventional implementations suffer from ``percussion smearing,'' a well-known artifact that significantly degrades the quality of percussive components. 
We attribute this artifact to a fundamental time-scale mismatch between the temporally smeared magnitude spectrogram and the localized, newly generated phase. 
To address this, we propose \mbox{SELEBI}, a signal-adaptive phase vocoder algorithm that significantly reduces percussion smearing while preserving stability and the perfect reconstruction property. 
Unlike conventional methods that rely on heuristic processing or component separation, our approach leverages the nonstationary Gabor transform. 
By dynamically adapting analysis window lengths to assign short windows to intervals containing significant energy associated with percussive components, we directly compute a temporally localized magnitude spectrogram from the time-domain signal. 
This approach ensures greater consistency between the temporal structures of the magnitude and phase. 
Furthermore, the perfect reconstruction property of the nonstationary Gabor transform guarantees stable, high-fidelity signal synthesis, in contrast to previous heuristic approaches. 
Experimental results demonstrate that the proposed method effectively mitigates percussion smearing and yields natural sound quality.
\end{abstract}


\section{Introduction}

\IEEEPARstart{T}{ime} stretching, a process that modifies the time-scale of a signal without altering its pitch, is a fundamental tool in modern music production, with applications ranging from audio remixing to transcription \cite{ishizaki2009full,risset2002examples}.
The ideal goal is to obtain a time-stretched signal that perfectly preserves the musical characteristics of the original, such as its timbre, clarity, and dynamics. 
To achieve high-quality results, a wide variety of time-stretching methods have been proposed \cite{flanagan1966phase,portnoff2003implementation,driedger2016review,verhelst1993overlap,moinet2011pvsola,doneRight,ottosen2017phase,driedger2013improving,nagel2009novel,duxbury2002improved,robel2003new,ravelli2005fast}.

\begin{figure}
    \centering
    \includegraphics[width=0.95\linewidth]{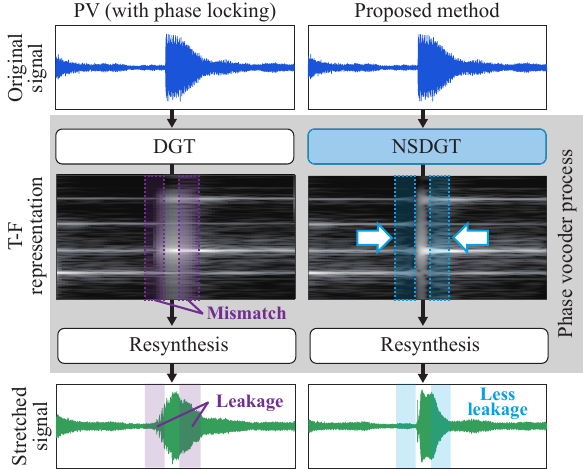}
    \caption{
    Block diagrams of the basic method, PV with identity phase locking \cite{laroche2002improved} (left), and the proposed method (right).
    By leveraging \mbox{NSDGT}, the proposed method synthesizes the target signal from T-F representations in which both the magnitude and phase spectrograms of percussive components are highly concentrated in the time direction.
    }
    \label{fig:propConcept}
\end{figure}

The phase vocoder (PV) \cite{flanagan1966phase,portnoff2003implementation} is one of the most widely used techniques for time-stretching. 
This approach operates in the time-frequency (T-F) domain: It computes the short-time Fourier transform (STFT), generates a new phase appropriate for the modified time-scale, and resynthesizes the signal using the overlap-add (OLA) technique with an extended hop size. 
Although the PV is a well-established technique that has been continuously refined \cite{driedger2016review,driedger2013improving}, its fundamental logic builds on the sinusoidal signal model, i.e., it assumes that the signal is a sum of sinusoids. This model is effective for tonal content because time-shifts are represented by a well-defined, predictable phase shift.
However, it is ill-suited for percussive components, which cannot be represented by a sum of few sinusoids. Therefore, time-shifting of percussive components cannot be modeled as such a phase shift. 
As a result, the PV often introduces a well-known artifact known as \textit{percussion smearing}.
As illustrated in Fig.\:\ref{fig:propConcept} (top-right), the stretched percussive components suffer significant degradation, losing their original temporal characteristics due to undesired leakage.

To alleviate this, various ``percussion-aware'' methods have been proposed. 
However, they either rely on an \textit{a priori} separation of the signal into tonal and percussive components for individual processing \cite{driedger2013improving,nagel2009novel} (Class A), which invariably introduces artifacts due to imperfect separation, or they suffer, to various degrees, from magnitude-phase mismatch~\cite{ottosen2017phase,duxbury2002improved,robel2003new,ravelli2005fast} (Class B), inhibiting the elimination of percussion smearing.

In this paper, we address the fundamental limitation of Class B methods\footnote{Our approach does not rely on signal separation, thereby avoiding the artifacts associated with Class A methods.
This allows us to focus solely on the limitations of Class B methods.}: the inconsistency between magnitude and phase. 
While these methods often succeed in preserving the phase relationships required for transients (i.e., vertical phase coherence), the corresponding magnitude spectrograms are inevitably smeared due to the time-stretching process. 
This results in percussive components being spread across a dilated time interval, which contradicts the localized phase information. 
We posit that this mismatch is the primary cause of percussion smearing, as illustrated in Fig.~\ref{fig:propConcept} (left). 
To address this, we propose computing spectrograms with improved temporal localization specifically where percussive components are present, thereby aligning the magnitude representation with the localized phase (Fig.~\ref{fig:propConcept}, right). 
While our previous work~\cite{akaishi2023} attempted to mitigate this issue through heuristic time-frequency bin shifting, the proposed method provides a theoretical foundation for energy preservation and stable synthesis.

To achieve this magnitude squeezing in a mathematically rigorous manner, this paper proposes a method named \textbf{\mbox{SELEBI}} (\underbar{\textbf{SELE}}ctive window compression with sta\underbar{\textbf{B}}le \underbar{\textbf{I}}nversion). 
Our method leverages the Nonstationary Gabor Transform (\mbox{NSDGT})~\cite{balazs2011theory}, a time-frequency representation framework that allows for adaptive, non-uniform windowing and sampling while retaining perfect reconstruction properties. 
We exploit this flexibility by adaptively assigning shorter windows to percussive regions, thereby directly obtaining a magnitude spectrogram with a desired temporal resolution. 
Crucially, the underlying mathematical framework of the \mbox{NSDGT} guarantees that this adaptive processing results in a stable and high-fidelity synthesis.

The rest of the paper is organized as follows.
Section II reviews the fundamentals of the \mbox{NSDGT} and PV-based time stretching. 
Section III discusses the core component of the proposed method: Sharply resolving percussive events to reduce percussion smearing in time-stretched audio. 
Section IV presents the proposed \mbox{SELEBI} algorithm. 
Section V discusses the feasibility of a bounded-delay implementation for on-line applications.
Section VI provides experimental results validating our method, and Section VI concludes this paper.

\section{Preliminaries}

\subsection{Notation}

The sets of natural numbers, integers, real numbers, and complex numbers are denoted by $\mathbb{N}$, $\mathbb{Z}$, $\mathbb{R}$, and $\mathbb{C}$, respectively.
Matrices are denoted by bold capital letters (e.g., $\mathbf{A}$), and their element at the $i$-th row and $j$-th column is denoted by $A[i,j]$.
Vectors are denoted by bold lower-case letters (e.g., $\mathbf{v}$), and their $i$-th element is denoted by $v[i]$.
For the purpose of this paper, we identify the cyclic group of order $N$, denoted by $\mathbb{Z}_{N}$, with the integers $0,\ldots,N-1$.
The following operations are always considered pointwise: $\odot$ (multiplication) $\oslash$ (division), $|\cdot|$ (modulus / absolute value), and $\text{Arg}(\cdot)$ (complex argument). 
A sequence indexed by $k \in \mathscr{K}\subset \mathbb{N}$ is denoted by $\{\cdot_k\}_{k \in \mathscr{K}}$.
The floor, ceiling, and nearest-integer (rounding) functions are denoted by $\lfloor\cdot\rfloor$, $\lceil\cdot\rceil$, and $\lfloor\cdot\rceil$, respectively.
The Frobenius norm is denoted by $\|\cdot\|_{\text{F}}$.

\subsection{Discrete Gabor Transform (DGT)}

The DGT coefficients $\mathbf{X}\in\mathbb{C}^{M\times N}$ of a signal $\mathbf{x}\in\mathbb{R}^{L}$ with a window function $\mathbf{g}\in\mathbb{R}^{L}$ are defined as \cite{grochenig}
\begin{equation}
\label{eq:DGT}
    X[m,n] = \sum^{L-1}_{l=0}x[l]\,g[l-an]\,\text{e}^{-\text{i}2\pi m(l-na)/M}, 
\end{equation}
where $\text{i}=\sqrt{-1}$, $a\in\mathbb{N}$ is the hop size, $l\in\mathbb{Z}_{L}$ is the time index, $n\in\mathbb{Z}_{N}$ is the time-frame index, $m\in\mathbb{Z}_{M}$ is the frequency index.
The signal length $L$ is set to satisfy $aN = bM = L$, where $b\in\mathbb{N}$ is the frequency decimation parameter. 

The inverse DGT (iDGT) is given by overlap-adding the individual, synthesized time frames: 
\begin{equation}
    \widehat{x}[l] = \sum_{n=0}^{N-1} \widetilde{g}[l-na]\cdot \sum_{m=0}^{M-1} X[m,n]\,\text{e}^{\text{i}2\pi ml/M},\label{eq:synth}
\end{equation}
where $\mathbf{\widetilde{g}}$ is the synthesis window. When the (analysis) window $\mathbf{g}$ has no more that $M$ (consecutive) nonzero samples, then by choosing \begin{equation}\label{eq:SynWin}\widetilde{g}[l] = \frac{g[l]}{M\sum_{n=0}^{N-1}g[l-na]^2},\end{equation}
we achieve error-free reconstruction, i.e., $x[l] = \widehat{x}[l]$, provided that the denominator in Eq.~\eqref{eq:SynWin} is strictly positive. 

As commonly done in the acoustics literature, we refer to $\mathbf{X}$ as \textit{(complex) spectrogram}.
Furthermore, $\mathbf{M} = |\mathbf{X}|$ and $\mathbf{\Phi} = \text{Arg}(\mathbf{X})$ are termed the \textit{magnitude} and \textit{phase spectrograms}, respectively \cite{yatabe2019representation}.

\begin{figure}[t]
    \centering
    \includegraphics[width=1\linewidth]{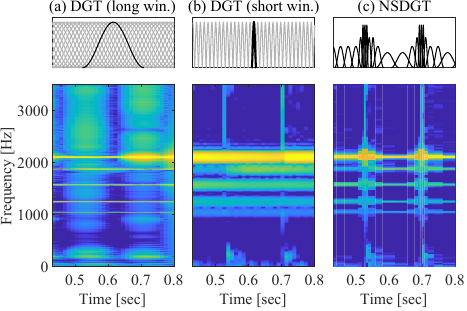}
    \caption{
    Comparison of windows and spectrograms for the DGT and \mbox{NSDGT}. 
    The upper boxes illustrate the window shift in the time domain, where representative DGT windows are highlighted in black for clarity. 
    The bottom boxes display the corresponding spectrograms for (a) DGT with a long window, (b) DGT with a short window, and (c) the \mbox{NSDGT}.
    }
    \label{fig:compareTF}
\end{figure}

\subsection{Nonstationary DGT (\mbox{NSDGT})}
\label{ssec:NSDGT}

The \mbox{NSDGT} \cite{balazs2011theory,dorfler2014nonstationary} generalizes the DGT by integrating variable hop sizes, window functions and numbers of frequency channels per time-frame into a flexible, invertible T-F representation. In this paper, however, we only consider variation of hop sizes and window functions, while fixing the number of frequency channels. 
For the $n$-th time frame, let $a_n \in \mathbb{N}$ be the hop size, $\mathbf{g}_n\in\mathbb{R}^{W_n}$ be the window function, $W_n\in\mathbb{N}$ be the window length, and $M \in \mathbb{N}$ be the number of frequency channels. Further define by $A_0 = 0$ and $A_n = \sum_{j=1}^n a_n$ the time position of the $n$-th time frame.
Hence, the \mbox{NSDGT} coefficients $\mathbf{X}^{\text{NS}}\in\mathbb{C}^{M\times N}$ as per \cite{balazs2011theory} are given by 
\begin{equation}
\label{eq:NSDGT}
    X^{\text{NS}}[m,n] = \sum^{L-1}_{l=0}x[l]\,g_{n}[l-A_n]\,\text{e}^{-\text{i}2\pi m(l-A_n)/M}.
\end{equation}
For $\mathbf{X}^{\text{NS}}$, $|\mathbf{X}^{\text{NS}}|$ and $\text{Arg}(\mathbf{X}^{\text{NS}})$ we adopt the terminology used for the DGT, i.e., we refer to them as \textit{(magnitude/phase) spectrogram}.

Fig.\:\ref{fig:compareTF} provides an example of a flexible T-F representation using the \mbox{NSDGT}.
As shown in (a) and (b), the standard DGT exhibits a well-known trade-off: a long window achieves good frequency localization for sinusoidal components, while a short window provides good temporal localization for percussive components. Most realistic signals, however, contain both component types, often even simultaneously. 
In contrast, the \mbox{NSDGT} (c) employs adaptive windowing with long windows for sinusoidal regions, short windows for percussive regions \cite{balazs2011theory}, and intermediate lengths for mixed components, resulting in improved time-frequency localization for the entire signal.

\subsection{PV-based Time Stretching}
\label{ssec:PV}

The PV begins by computing a DGT, see Eq.~\eqref{eq:DGT}, with the analysis hop size $a$.
Subsequently, the phase spectrogram is modified while leaving the magnitude spectrogram unchanged.
Finally, a time-stretched signal is synthesized by applying an inverse DGT (iDGT) with the synthesis hop size $\widetilde{a}$ given by $\widetilde{a} = \lceil\alpha a\rceil$.
Here, $\alpha\in\mathbb{R}_{+}$ is the desired stretching factor\footnote{Percussion smearing is observed only when the time-scale is extended, such that we consider $\alpha >1$ here.}. This results in the final time-stretched signal $\widehat{\mathbf{x}}\in\mathbb{R}^{\widetilde{a}N}$.

Restricting, for the sake of a more concise treatment, to the case of constant stretch factor $\alpha$, the main difference between PV variants concern the specific modification applied to the phase spectrogram. In the classical phase vocoder~\cite{flanagan1966phase,portnoff2003implementation}, the new phase spectrogram $\mathbf{\widetilde{\Phi}}$ is computed by scaling the time-direction partial derivative $\mathbf{\Delta_{\text{t}}\Phi}$ of the phase $\mathbf{\Phi}$ with $\alpha$ before integrating the phase along time within each channel. Commonly, this process is performed in two steps: The derivative is approximated as
\begin{equation}
\label{eq:HPD}
\begin{split}
    (\Delta_{\text{t}}&\Phi)[m,n]=\\
    \frac{1}{a}&\left[\Phi[m,n]-\Phi[m,n-1]-\frac{2\pi ma}{M}\right]_{2\pi}+\frac{2\pi m}{M},
\end{split}
\end{equation}
where $\left[\cdot\right]_{2\pi}=\cdot-2\pi\lfloor\cdot/2\pi\rceil$ is the principal argument calculation. Subsequently, $\mathbf{\widetilde{\Phi}}$ is computed with the recursive phase propagation formula \cite{laroche2002improved}, 
\begin{equation}
\label{eq:RPPformula}
    \widetilde{\Phi}[m,n] = \widetilde{\Phi}[m,n-1]+\widetilde{a}\,(\Delta_{\text{t}}\Phi)[m,n].
\end{equation}
Various heuristic modifications have been proposed to the estimation of $\mathbf{\Delta_{\text{t}}\Phi}$ or the integration step, in order to improve perceptual quality~\cite{laroche2002improved,puckette1995phase}, some of which have been integrated in a prior extension of the PV using \mbox{NSDGT}~\cite{ottosen2017phase}.

Conventional phase generation in PV-based time stretching typically relies solely on the time-direction phase derivative to model phase evolution, as in Eq.~\eqref{eq:HPD} and \eqref{eq:RPPformula}. 
This approach essentially assumes a sinusoidal model, treating frequency channels independently. 
Consequently, it disregards the vertical phase relationships that are crucial for preserving transient sharpness, leading to percussion smearing~\cite{driedger2013improving}. 
To mitigate this, it is common to apply \emph{phase-locking} techniques~\cite{ottosen2017phase,laroche2002improved,puckette1995phase,laroche1997phase}, which heuristically enforce consistency between adjacent channels to maintain vertical coherence.

Moving beyond heuristics, the method proposed in \cite{doneRight} introduces a more rigorous framework by adapting \emph{Phase Gradient Heap Integration}~\cite{pruvsa2016real}, a technique originally developed for phaseless reconstruction. 
Unlike standard PV or simple phase-locking, this approach considers the full phase gradient; that is, it estimates and scales both the time-direction derivative $\mathbf{\Delta_{\text{t}}\Phi}$ and the frequency-direction derivative $\mathbf{\Delta_{\text{f}}\Phi}$. 
By performing adaptive numerical integration along the optimal path in the time-frequency plane, this method achieves significant improvements in perceptual quality and reduces percussion smearing without relying on potentially unreliable heuristics.

However, even with advanced phase generation techniques like \cite{doneRight} or other percussion-aware phase refinements~\cite{ottosen2017phase,duxbury2002improved,robel2003new,ravelli2005fast}, artifacts remain unavoidable, especially at large stretching factors $\alpha$. 
This is due to a fundamental mismatch between the estimated phase and the magnitude spectrogram. 
While these methods strive to reconstruct a temporally localized phase corresponding to a transient, the underlying magnitude spectrogram $\mathbf{M}$ remains temporally smeared due to the stretching process. 
This inconsistency between the localized phase and the spread magnitude prevents the complete elimination of percussion smearing.

\section{Reducing Phase-Magnitude Mismatch Through Improved Transient Concentration}
\label{sec:TDS}

Our proposed method is based on the hypothesis that, besides imperfect phase estimates, the primary cause of percussion smearing is a time-scale mismatch between the magnitude $\mathbf{M}$ and the estimated phase $\mathbf{\widetilde{\Phi}}$ used to synthesize the time-stretched signal. 
Successful improvements to the PV, such as \cite{robel2003new,doneRight}, use the magnitude as a cue to generate a localized phase, implicitly or even explicitly enforcing a stretch factor of $1$ for percussive events. 
However, the unchanged magnitude spectrogram, when interpreted using the synthesis hop size $\widetilde{a}$, indicates temporal smearing of said events, leading to artificial lengthening during synthesis, which cannot be entirely suppressed by a better phase estimate. 
To overcome this, our key idea is to construct a magnitude spectrogram that sharply represents percussive events. 
We employ a dual strategy: (1) shortening the analysis window to enhance temporal resolution, and (2) reducing the number of time-frames covering the event to minimize inter-frame smearing. 
By allowing these two mechanisms to work in tandem, we aim to keep the transient representation sharp, effectively preserving its original time-scale, while allowing the tonal resonances to be stretched naturally.

\begin{figure}
    \centering
    \includegraphics[width=1\linewidth]{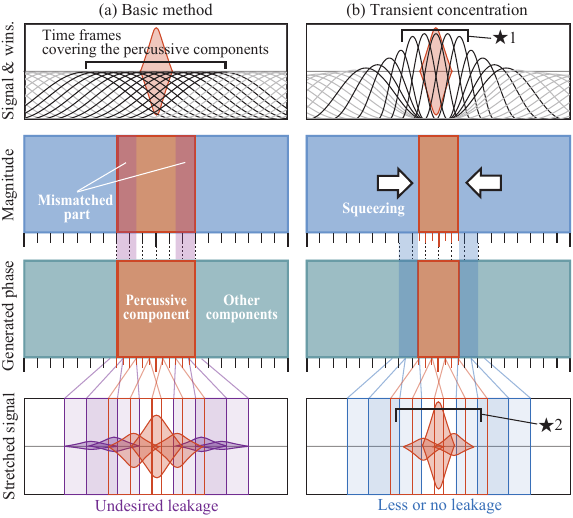}
    \caption{
    Conceptual illustration of time-directional spectrogram ``squeezing.'' (a) The conventional PV-based method using DGT. (b) The proposed method utilizing transient concentration. 
    The top row displays the input time-domain signal (amplitude vs.\ time) and the analysis window functions. 
    The second and third rows schematically represent the magnitude spectrogram and the corresponding generated phase, respectively. 
    The bottom row shows the synthesized time-stretched signal.
    In these schematic representations, the percussive component is colored red, and the windows capturing this component are emphasized (non-percussive components are omitted in this row for clarity).
    In the spectrograms, the red area highlights the percussive component, while the blue and green areas represent the magnitude and phase of the other components, respectively.
    Because the percussive interval is maintained at its original time-scale, it appears compressed relative to the new, stretched time axis (illustrated below the spectrograms). 
    The bottom panel details the synthesis of the percussive component across the new time frames.
    The markers $\star 1$ and $\star 2$ highlight the key innovations of the proposed method: shortening the window length and reducing the number of time frames, respectively.
    }
    \label{fig:TDS}
\end{figure}

Fig.\:\ref{fig:TDS} illustrates the effect of this magnitude ``squeezing''.
In the standard PV approach (Fig.\:\ref{fig:TDS}(a)), the use of a fixed, long analysis window with high overlap distributes the percussive energy across a broad sequence of time frames. 
When stretched, this sequence expands, inevitably causing the synthesized transient to be less sharp and introducing phase-magnitude mismatch (purple regions).
In contrast, our method (Fig.\:\ref{fig:TDS}(b)) dynamically adapts the representation. 
By simultaneously shortening the window length ($\star 1$) and reducing the frame density ($\star 2$) in the vicinity of the transient, the method effectively ``squeezes'' the signal energy into a narrower region on the stretched grid. 
This concentration ensures that the transient is synthesized with a sharpness comparable to the original signal, drastically reducing leakage and smearing.

In our previous conference paper \cite{akaishi2023}, we presented a preliminary, heuristic method designed to achieve a similar effect. In that method, we shift T-F bins horizontally toward the center of the corresponding percussive event. For compatibility with the used DGT and existing phase estimation schemes for time-stretching, e.g., Eq.~\eqref{eq:RPPformula}, the resulting, non-uniform spectrogram is resampled on a uniform temporal grid after spline interpolation. 
Although this approach provides an acceptable improvement in percussive sound quality, its heuristic design introduces several potential issues: 1) The bin-shifting operation is not energy-preserving; temporally localizing the bins without properly concentrating their energy led to a slight reduction in volume. 2) Spline interpolation of the spectrogram neglects the effects of original choice of analysis window and hop size, introducing approximation errors.

\section{Proposed Method: \mbox{SELEBI}}
\label{sec:prop}

In this section, we introduce  \mbox{SELEBI}, the proposed time-stretching method designed to suppress percussion smearing. 
As discussed in the previous section, this is achieved by designing a T-F representation that enforces temporal concentration of transients. 
To this end, we employ the \mbox{NSDGT}, locally adapting both the window lengths $W_n$, as seen in Fig.\:\ref{fig:TDS}(b), and hop sizes $a_n$, in a small neighborhood of detected percussive events.

The efficacy of this approach is demonstrated in Fig.\:\ref{fig:TDS}. While a standard DGT magnitude spectrogram temporally smears an impulse (second row, left), our \mbox{NSDGT}-based spectrogram maintains sharp localization (second row, right).
Crucially, the \mbox{NSDGT} framework resolves the limitations of our previous method \cite{akaishi2023}: Unlike the simple bin-shifting which resulted in energy loss, the \mbox{NSDGT} is mathematically rigorous and invertible, thereby guaranteeing energy preservation. 
Furthermore, in contrast to the interpolated spectrogram in \cite{akaishi2023}, the \mbox{NSDGT} structure retains well-defined window and hop parameters, leading to a better phase estimate.

\subsection{Flow of \mbox{SELEBI}}
\label{ssec:flow}

\begin{figure}
    \centering
    \includegraphics[width=0.9\linewidth]{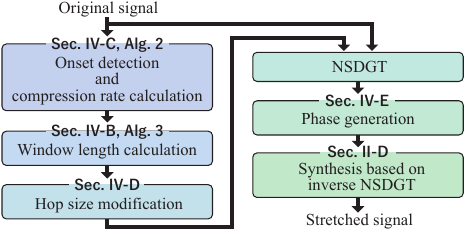}
    \caption{
    The flow of the proposed method. 
    The left column details the \mbox{NSDGT} parameter calculation, while the right column illustrates the subsequent processing steps.
    }
    \label{fig:flow}
\end{figure}

The flow of \mbox{SELEBI} is summarized in Alg.\:\ref{alg:proc} and illustrated in Fig.\:\ref{fig:flow}. 
Our method first performs a preprocessing stage (lines 1--3) to determine the \mbox{NSDGT} parameters, including window lengths and hop sizes. 
To appropriately modify the windows, the algorithm first identifies the position of percussive components (line 1, Alg.\:\ref{alg:compRatio}), using onset detection based on 
harmonic-percussive sound separation (HPSS) \cite{fitzgerald2010harmonic}. 
It then calculates a corresponding \textit{compression rate}, defined as the proportion of frequency bins in that frame that are classified as percussive. The compression rate determines the shortest window length, used at the onset position. This procedure ensures that strong transient events remain highly concentrated and reduces the presence of audible artifacts when transient and harmonic events overlap. 

A preliminary (constant) analysis hop size is set based on the selected time-scale factor, such that, under the assumption that the longest admissible window length be used everywhere, sufficient overlap remains after time-stretching. At each time position, a desired window length is determined based on the detected onsets and compression rates (line 2, Alg.\:\ref{alg:winlenCalc}). In regions where the window length is reduced, the original hop size is too large and would lead to insufficient coverage, such that new, narrower time positions are adaptively chosen for appropriate coverage, while ensuring that the number of time frames covering the percussive event remains small (line 3, Alg.\:\ref{alg:winlenCalc}). 
Using these parameters, the \mbox{NSDGT} is computed to obtain a ``squeezed'' magnitude spectrogram (line 4). 
Subsequently, a modified phase generation procedure, adapted to be compatible with variable hop sizes, is applied (line 5), and the time-stretched signal is synthesized using the inverse \mbox{NSDGT} (line 6).

The following subsections detail each part of this process: parameter calculation (Sec.\:\ref{ssec:compressionRate}), selective window compression (Sec.\:\ref{ssec:winLenMethod}), hop size determination (Sec.\:\ref{ssec:light}), and modified phase generation (Sec.\:\ref{ssec:phase}).
Although the proposed method uses an adaptive hop size, the figures in Sec.\:\ref{ssec:compressionRate} and Sec.\:\ref{ssec:winLenMethod} show a fixed hop size for simplicity.

\begin{algorithm}[t]
{\footnotesize
\caption{Proposed method (\mbox{SELEBI})}
\label{alg:proc}
\begin{algorithmic}[1]
\renewcommand{\algorithmicrequire}{\textbf{Input:}}
\renewcommand{\algorithmicensure}{\textbf{Output:}}
\Require $\mathbf{x}\in\mathbb{R}^{L}$
\Ensure $\widehat{\mathbf{x}}\in\mathbb{R}^{\alpha L}$
\State $\text{paramsForWinLen}=\text{computCompRate}(\mathbf{x})$\Comment{Sec.\:\ref{ssec:compressionRate}, Alg.\:\ref{alg:compRatio}}
\State $\text{winLen}=\text{computWinLen}(\text{paramsForWinLen})$\Comment{Sec.\:\ref{ssec:winLenMethod}, Alg.\:\ref{alg:winlenCalc}}
\State $[\text{winLen},\text{hopSize}]=\text{modifyHopSize}(\text{winLen})$\Comment{Sec.\:\ref{ssec:light}}
\State $\mathbf{X}^{\text{NS}}=\text{NSDGT}(\mathbf{x},\text{winLen},\text{hopSize})$\Comment{Eq.~\eqref{eq:NSDGT} in Sec.\:\ref{ssec:NSDGT}}
\State $\mathbf{\Phi}=\text{genPhase}(\mathbf{X}^{\text{NS}},\text{hopSize})$\Comment{Sec.\:\ref{ssec:phase}}
\State $\widehat{\mathbf{x}}=\text{synthesis}(\mathbf{X}^{\text{NS}},\mathbf{\Phi},\text{hopSize})$\Comment{Sec.\:\ref{ssec:PV}}
\end{algorithmic}
}
\end{algorithm}

\begin{algorithm}[t]
{\footnotesize
\caption{Computation of the compression rate}
\label{alg:compRatio}
\begin{algorithmic}[1]
\renewcommand{\algorithmicrequire}{\textbf{Input:}}
\renewcommand{\algorithmicensure}{\textbf{Output:}}
\Require $\mathbf{x}\in\mathbb{R}^{L}$
\Ensure $\{r_k\}_{k\in\{1,\dots,K\}}, \{I_k\}_{k\in\{1,\dots,K\}}$
\State $\mathbf{X} = \text{DGT}(\mathbf{x})$\Comment{compute the spectrogram}
\State $\mathbf{\Phi}_{\text{m}} = \partial^2\text{Arg}(\mathbf{X})/\partial \tau\partial\omega$\Comment{compute the MPD of phase}
\State $\mathbf{M}=\mathcal{M}(\mathbf{X},\mathbf{\Phi}_{\text{m}})$
\State\Comment{compute the mask for separating percussive components}
\State $\mathbf{X}_{\text{p}} \,= \mathbf{X}\odot \mathbf{M}$\Comment{compute the masked spectrogram}
\State $\mathbf{r} = \text{filtering}(\sum_{m=0}^{M-1}|X_{\text{p}}[m,n]|\oslash\sum_{m=0}^{M-1}|X[m,n]|)$
\State\Comment{compute the ratio of the percussive components in each frame}
\State $\left[\{r_k\}_{k\in\{1,\dots,K\}}, \{I_k\}_{k\in\{1,\dots,K\}}\right] = \text{findPeaks}(\mathbf{r})$
\State\Comment{find the positions of the pulse and decide their compression rates}
\end{algorithmic}
}
\end{algorithm}

\begin{algorithm}[t]
{\footnotesize
\caption{Computation of the window length vector}
\label{alg:winlenCalc}
\begin{algorithmic}[1]
\renewcommand{\algorithmicrequire}{\textbf{Input:}}
\renewcommand{\algorithmicensure}{\textbf{Output:}}
\Require $\{r_k\}_{k\in\{1,\dots,K\}}, \{I_k\}_{k\in\{1,\dots,K\}}$, $N_{\text{half}}=\lceil V/(2a)\rceil$
\Ensure $\mathbf{v}\in\mathbb{N}^N$
\For{$k = 1,\dots,K$}\Comment{for all percussive components}
\State $S_k =\lfloor V - r^{2}_{k}\,(1-1/\alpha)\,V\rfloor$
\State\Comment{calculate the smallest window length}
\For{$n = 1,\dots,N$}\Comment{for all time indices}
\If{$|n - I_k|\le N_{\text{half}}$}\Comment{within the percussive intervals}
\State $V[k,n] = \text{max}(S_k,2a|n - I_k| + V - 2aN_{\text{half}})$
\State\Comment{shorten the window length}
\Else
\State $V[k,n] = V$\Comment{retain the length}
\EndIf
\EndFor
\EndFor
\State $\mathbf{v} = \texttt{min}(V[1,:],\dots,V[K,:])$\Comment{select shorter lengths}
\end{algorithmic}
}
\end{algorithm}

\subsection{Onset Detection and Compression Rate Calculation}
\label{ssec:compressionRate}

In order to determine the adaptive window lengths used in \mbox{SELEBI}, it is necessary to identify the positions of percussive components and determine the corresponding compression rates. The window length compression itself is described in Sec.\:\ref{ssec:winLenMethod}. 
Our method is derived from the spectral flux and weighted phase deviation onset detectors described in \cite{dixon2006onset}. We combine these methods and extend them to quantify the relative intensity of percussive components at the detected onsets. As such, determining the \mbox{SELEBI} parameters involves computing the DGT of the original signal, classifying T-F bins as percussive or not, and finally quantifying the proportion of percussive content within each onset frame.

\begin{figure}[t]
    \centering
    \includegraphics[width=1\linewidth]{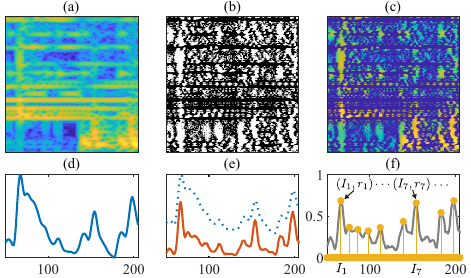}
    \caption{
    Example of the computation of the compression rate.
    From top left to bottom right, (a) the magnitude spectrogram $|\mathbf{X}|$, (b) the enhancement mask, (c) the enhanced spectrogram $|\mathbf{X}_{\text{p}}|$, (d) the frequency-directional sum of $|\mathbf{X}|$, (e) the frequency-directional sum of $|\mathbf{X}_{\text{p}}|$, and (f) the compression rate $\mathbf{r}$ (the detected peaks are plotted in yellow).
    The mask in (b) is colored in white where $\mathcal{M}(\mathbf{X},\mathbf{\Phi}_{\textrm{mix}})[m,n] = 1$.
    }
    \label{fig:compressionRate}
\end{figure}

Fig.\:\ref{fig:compressionRate} illustrates the principles applied for onset detection and compression rate selection by ways of an example. The magnitude ratio (red line in panel (e)) is computed for each time frame by dividing the sum of percussive T-F bins across frequencies by the total magnitude summed across frequencies (dotted line). The final parameters are then extracted by detecting the peaks of this ratio (panel (f))%
\footnote{For the detection of peaks of the ratio, we use the MATLAB function $\texttt{findpeaks}$ with ``$\texttt{MinPeakProminence}$'' $=0.1$.}, where the height and position of each peak correspond to the compression ratio $r_k$ and center position $I_k$ of a pulse. 
Alg.\:\ref{alg:compRatio} details this procedure. 
Note that local fluctuations in the magnitude ratios are suppressed by applying a one-dimensional median filter (line 6, Alg.\:\ref{alg:compRatio}). 
We now proceed to discuss the onset detection algorithm and the function $\mathcal{M}$ used to classify T-F bins as percussive, resulting in a binary mask $\mathbf{M}$.

\begin{figure*}[t]
    \centering
    \includegraphics[width=1\linewidth]{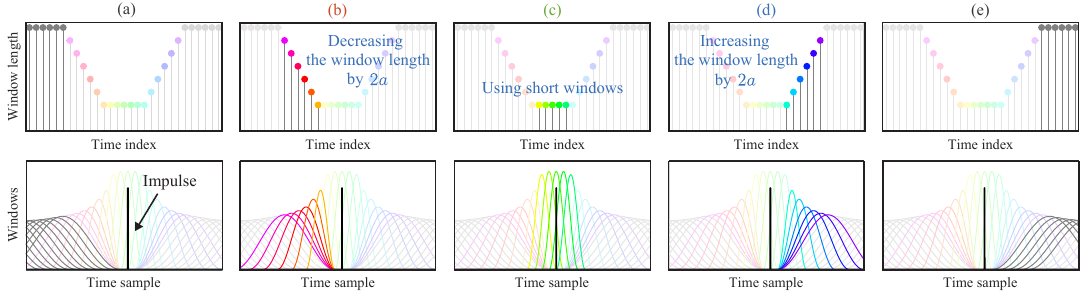}
    \caption{
    Concept of the window length adjustment in the proposed method. The top row shows the target window length, while the bottom row shows the corresponding analysis windows. The colors in the top row match those in the bottom row. Stages (a)--(e) illustrate the window behavior as it encounters an impulse.
    }
    \label{fig:winLen}
\end{figure*}

\begin{figure}[t]
    \centering
    \includegraphics[width=1\linewidth]{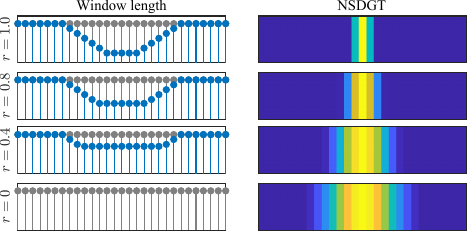}
    \caption{
    Effect of compression rate $r$ on window length (left) and the resulting impulse magnitude spectrogram (right). 
    The left panels plot the window lengths (vertical axis) versus time (horizontal axis), where the original window length is shown in gray, and the modified window length is shown in blue.
    The right panels show the corresponding magnitude spectrograms (frequency vs. time) of an impulse, obtained using the window functions from the left.
    From top to bottom, the rows show compression rates of $r = 1.0, 0.8, 0.4, \text{and } 0$, chosen for visibility.
    }
    \label{fig:changeRatio}
\end{figure}

To compute $\mathcal{M}$, we utilize the mixed-partial derivative (MPD) of the phase, $\text{MPD}(\omega,\tau) = (\partial^2\text{Arg}(Y)/\partial \tau\partial\omega)(\omega,\tau)$ as in \cite{akaishi2022harmonic}, whereas a threshold on the magnitude $|\mathbf{X}|$ is additionally used to ignore T-F regions with negligible energy, in which the phase is unstable. 
Fundamentally, an onset detection function based solely on $|\mathbf{X}|$ such as $(\partial |X|/\partial \tau)(\omega,\tau)$ could be used, we found that magnitude information alone is often insufficient for reliable onset detection in complex, real-world signals containing diverse components. In these settings, we obtained improved identification accuracy with MPD. For details on the suitability of MPD for classification of predominantly percussive T-F bins, see \cite{akaishi2022harmonic} and  \cite{fulop2007separation}, which argue that $\text{MPD}(\omega,\tau) \approx 0$ at T-F bins containing purely sinusoidal components and $\text{MPD}(\omega,\tau) \approx 1$ (possibly after scaling), at T-F bins containing only impulsive components, or 
\begin{equation}
    \!\!\!\frac{\partial^2}{\partial\tau\partial\omega}\text{Arg}(Y)(\omega,\tau)\approx\left\{
    \begin{array}{cl}
         0&\text{(if $Y(\omega,\tau)$ is sinusoidal)}, \\
         1&\text{(if $Y(\omega,\tau)$ is impulsive)}.
    \end{array}\right.
\end{equation}

Using these properties of MPD, we can create a binary mask that removes T-F bins that are not classified as percussive. 
For this paper, we consider the operator $\mathcal{M}:(\mathbb{C}^{M\times N},\mathbb{R}^{M\times N})\to\{0,1\}^{M\times N}$ for generating said mask, defined by 
\begin{equation}
    \mathcal{M}(\mathbf{X},\mathbf{\Phi}_{\text{mix}}) = \mathcal{M}_{\text{mag}}(\mathbf{X})\odot\mathcal{M}_{\text{p}}(\mathbf{\Phi}_{\text{mix}}),
\end{equation}
where $\mathbf{\Phi}_{\text{mix}}\in\mathbb{R}^{M\times N}$ is MPD corresponding to $\mathbf{X}$, 
\begin{equation}
\begin{array}{rl}
\mathcal{M}_{\text{mag}}(\mathbf{X})[m,n] =&\!\!\!\! \left\{ 
\begin{array}{cl}
1 & \text{if}\quad|X[m,n]| > \theta_{\text{mag}}, \\
0 & \text{otherwise},
\end{array}\right. \\
\mathcal{M}_{\text{p}}(\mathbf{\Phi}_{\text{mix}})[m,n] =&\!\!\!\! \left\{ 
\begin{array}{cl}
1 & \text{if}\quad\underline{\theta_{\text{p}}}<\Phi_{\text{mix}}[m,n]-1 < \overline{\theta_{\text{p}}}, \\
0 & \text{otherwise},
\end{array}\right.
\end{array}
\end{equation}
and $\theta_{\text{mag}}, \underline{\theta_{\text{p}}},\overline{\theta_{\text{p}}}>0$ are hyperparameters used for thresholding.
The two masks serve distinct purposes: $\mathcal{M}_{\text{mag}}$ eliminates low-magnitude (noisy) components, and $\mathcal{M}_{\text{p}}$ enhances percussive components.
Fig.\:\ref{fig:compressionRate}(b) shows the resulting mask $\mathbf{M} = \mathcal{M}(\mathbf{X},\mathbf{\Phi}_{\text{m}})$. This mask successfully reveals even subtle percussive components (highlighted in white), which leads to the effective enhancement of these components, as shown in (c).

\subsection{Selective Window Length Compression}
\label{ssec:winLenMethod}

Once the center position $I_k$, and the compression ratio $r_k\in[0,1]$ of percussive events has been determined, \mbox{SELEBI} adjusts the length of the analysis windows used in the neighborhood of $I_k$, such that temporal smearing is minimized. 
The adaptation process is illustrated in Fig.\:\ref{fig:winLen}. 
It ensures that the impulse is ``observed'' only by few, short windows, rather than being captured by many long windows that produce smearing after time-stretching. 
The length of the windows used is determined by the compression rate $r_k$, with larger $r_k$ enforcing shorter windows of length $S_k$ at and around the center of the $k$-th percussive event, as shown in Fig.\:\ref{fig:changeRatio}.
The length $S_k$ is determined by the formula: 
\begin{equation}
\label{eq:minWin}
    S_k =\lfloor V - r_k^2\,(1-1/\alpha)\,V\rfloor,
\end{equation}
where $V$ is the length of the original, long analysis window, and $\alpha>0$ is the stretching factor.
Note that the window length remains unchanged when $r_k=0$, as shown in the bottom row of Fig.\:\ref{fig:changeRatio}.
To ensure a stable representation of the signal surrounding the percussive event, \mbox{SELEBI} gradually varies the window length from the full length $V$ to the target length $S_k$ (and vice versa) by successively adjusting the lengths by multiples of the hop size (e.g., $2a$), as shown in Fig.\:\ref{fig:winLen}, panels (b) and (d).

We now summarize the procedure of calculating the window length for each time frame, see Alg.\:\ref{alg:winlenCalc} for a formal description. 
We begin by looking at each percussive event individually. 
Let $N_{\text{half}} = \lceil V/(2a)\rceil$ denote the number of frames between the edge of the (long) analysis window and its center.
This number determines the time indices left and right of $I_k$ that will be assigned an adapted window length. 
Starting from time index $J_{k,-} = I_k- N_{\text{half}}$, the window length is shortened by $2a$, until a length of $S_k$ is reached, i.e., at index $J_{k,-} + j$, we assign window length $\max\{V-2a(j-1),S_k\}$, for $0\leq j \leq N_{\text{half}}+1$. Likewise, with $J_{k,+} = I_k+ N_{\text{half}}$, we assign window length $\max\{V-2a(j-1),S_k\}$ to index index $J_{k,+} - j$, for $0\leq j \leq N_{\text{half}}+1$. This ensures that the center of the percussive event is covered by no more than $2\lceil S_k/(2a)\rceil+1$ windows of length $S_k$ and all other analysis time frames do not contain said center. This process is illustrated in Fig.\:\ref{fig:winLen}.

\begin{figure}
    \centering
    \includegraphics[width=0.75\linewidth]{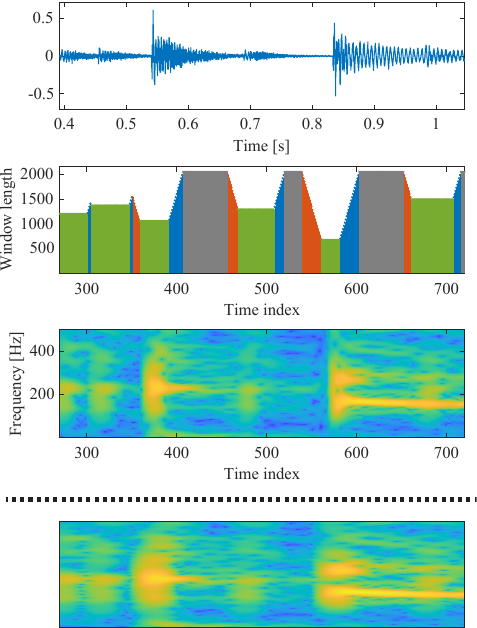}
    \caption{
    Example of adaptive window length adjustment and the resulting \mbox{NSDGT} spectrogram for a bongo signal. 
    The panels display, from top to bottom: the original waveform, the adaptive window length at each time index, the resulting \mbox{NSDGT} spectrogram, and the standard DGT spectrogram.
    For clarity, the middle panel is color-coded to correspond with the adjustment stages shown in Fig.\:\ref{fig:winLen}: non-percussive regions (gray; (a), (e)), pre-pulse (red; (b)), pulse center (green; (c)), and post-pulse (blue; (d)).
    }
    \label{fig:example1}
\end{figure}

In the case that the intervals $[J_{k,-},J_{k,+}]$ overlap, for different values of $k$, we select the minimum of all assigned window lengths. This can be seen in the example shown in Fig.\:\ref{fig:example1}.

\subsection{Modification of Hop Sizes}
\label{ssec:light}

In practice, using varying window sizes at equidistant time positions $na$ is either computationally wasteful (small hop size $a$) or leads to synthesis instability where short windows are used (large hop size $a$).
Whereas the former has negative impact on the algorithm's demands in terms of computation time and hardware, the latter may produce synthesis artifacts and lead to poor quality audio output. 
To resolve this, we construct a variable hop size \mbox{NSDGT} by choosing hop sizes adapted to the window length determined in Sec.\:\ref{ssec:winLenMethod}. 
Specifically, we employ a standard analysis hop size, $a$, for non-percussive regions (long windows) and an adaptive hop size, $\widehat{a}_{k}$, for percussive regions (short windows).
In the transition regions, we smoothly vary the hop sizes and window lengths via interpolation to prevent sampling artifacts.
To maintain sufficient overlap in the percussive regions, $\widehat{a}_{k}$ is derived from the short window length $S_k$ and the time-stretching factor $\alpha$, which will be defined later.

We describe the hop size determination process.
First, the time frames are segmented into four distinct region types based on the target window length configuration: (i) constant (long), (ii) constant (short), (iii) transition (long to short), and (iv) transition (short to long).
The time positions and window lengths at the boundaries of these regions are fixed to the original grid.
For type (i) regions, the original grid remains unaltered.
In type (ii) regions, we introduce an adaptive hop size defined as $\widehat{a}_{k} = \lfloor S_{k}/(\alpha\beta)\rceil$, where $\beta > 1$ is a parameter controlling the overlap ratio (empirically set to $\beta=4$).
To fill the fixed region duration $N_{\text{org}} a$, where $N_{\text{org}}$ is the number of frames in the original grid, we assign the hop size $\widehat{a}_{k}$ sequentially from the beginning of the region.
The final hop size serves as a compensation frame to absorb the residual time difference\footnote{If the duration is divisible by $\widehat{a}_{k}$ without remainder, this residual is zero.}, calculated as $a N_{\text{org}} - \widehat{a}_{k} \lfloor a N_{\text{org}}/\widehat{a}_{k} \rfloor$.
In cases where two percussive events occur in close proximity, type (ii) regions may become adjacent; in such instances, the window length and hop size change stepwise between these regions.

For transition regions (types (iii) and (iv)), the hop size is varied linearly to smoothly connect the adjacent regions.
To determine the linear trajectory, the starting and ending hop sizes, denoted as $a_{\text{st}}$ and $a_{\text{end}}$, must first be defined based on the transition pattern summarized in Table~\ref{tab:hopsizeDefinition}.
Patterns (A) and (B) simply utilize the standard hop size $a$ and the adaptive hop size $\widehat{a}_{k}$.
Conversely, patterns (C) and (D) represent cases where the window length trajectory switches direction (i.e., it expands and immediately transitions to shrinking) without returning to the long window state (type (i)), as seen around the 350-th sample in Fig.\:\ref{fig:example1}.
In such cases, we define an intermediate hop size based on the boundary window length $V_{\text{bnd}}$ as $\breve{a}_{k} = \lfloor a\,V_{\text{bnd}}/V \rceil$.
According to the window length formulation in Sec.\:IV-C, $V_{\text{bnd}}$ between the $k$-th and $(k+1)$-th percussive components is determined as $V_{\text{bnd}} = a(I_{k+1}-I_{k})+V-2aN_{\text{half}}$.

We state the procedure focusing on type (iii) transitions.
To maintain the total duration of the region, $N_\text{org} a$, while linearly varying the hop size, we first calculate the target number of new frames, $N_{\text{new}}$, as
\begin{equation}
N_{\text{new}} = \left\lfloor \frac{2}{a_{\text{st}} + a_{\text{end}}}\left(a N_{\text{org}} - \frac{a_{\text{st}} - a_{\text{end}}}{2}\right)\right\rfloor.
\end{equation}
To avoid generating unnaturally small compensation hop sizes in regions where larger hop sizes are intended, we arrange the hop positions starting from the boundary with the larger window length.
Consequently, the hop size $a_l$ at the $l$-th frame from this starting point is defined as
\begin{equation}
a_l = \left\lfloor a_{\text{max}} - \frac{l}{N_\text{new}} (a_{\text{max}} - a_{\text{min}})\right\rfloor,
\end{equation}
where $a_{\text{max}}$ and $a_{\text{min}}$ denote the larger and smaller values between $a_{\text{st}}$ and $a_{\text{end}}$, respectively.
This formulation is derived by solving $\sum_{l=1}^{N_{\text{new}}} a_l = N_\text{org} a$ under an integer constraint.
Any residual time difference due to rounding is compensated for by inserting a correction hop size of $\lfloor(N_{\text{org}} a - \sum_{l=1}^{N_{\text{new}}} a_l)/2\rfloor$ at the end of the sequence (i.e., near the short windows).
Type (iv) transitions are treated analogously, starting from the endpoint corresponding to the larger window length (i.e., the end of the region).
Once the hop sizes are determined, the window lengths for the new grid are derived via linear interpolation based on the correspondence with the original grid.

\begin{figure}[t]
    \centering
    \includegraphics[width=1\linewidth]{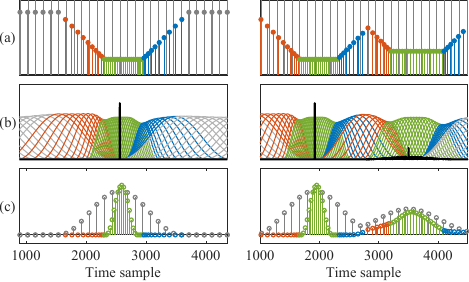}
    \caption{
    Example of smooth hop size adjustment and impulse observation (corresponding to Sec.\:\ref{ssec:light}). 
    The left column shows a single impulse, while the right column shows two adjacent impulses of differing amplitudes.
    Note that the small impulse on the right is embedded in Gaussian noise to simulate a less-percussive sound. 
    (a) plots the adaptive window length (stems) at each hop position, with gray line indicating the positions of a large, fixed hop size; (b) illustrates the corresponding shifting windows; and (c) shows the observed amplitude, with the DGT-based magnitude in gray for comparison. 
    The window coloring scheme is the same as in Fig.\:\ref{fig:example1}.
    }
    \label{fig:changeHopSize}
\end{figure}

An example of the hop size modification is illustrated in Fig.\:\ref{fig:changeHopSize}.
The timeline is categorized into three types of regions based on window length behavior: decreasing (red), constant (green), and increasing (blue).
The constant regions correspond to either the non-percussive state (using $a$) or the percussive state (using $\widehat{a}$).
In the transition regions, the hop size varies linearly.
This strategy effectively handles adjacent percussive events, as shown in Fig.\:\ref{fig:changeHopSize} (right): the hop size widens after the first pulse (approx.\ 2500--2800-th sample) before narrowing again for the second.
This adaptive sampling ensures that the ``window compression'' described in Sec.\:\ref{ssec:winLenMethod}, intended to limit transient observation, is successfully achieved.

\begin{table}[t]
    \centering
    \caption{Hop size definitions for each transition type.}
    \label{tab:hopsizeDefinition}
    \begin{tabularx}{\columnwidth}{cYYYYY}
        \toprule
        & \multicolumn{3}{c}{Type of transitions} & \multicolumn{2}{c}{Hop size settings} \\
        \cmidrule(lr){2-4} \cmidrule(lr){5-6}
        & Before & Transition & After & $a_{\text{st}}$ & $a_{\text{end}}$ \\
        \midrule
        (A) & \textcolor[HTML]{505050}{(i)}  & \textcolor[HTML]{d95319}{(iii)}       & \textcolor[HTML]{77ac30}{(ii)} & $a$ & {$\widehat{a}_{k}$} \\
        (B) & \textcolor[HTML]{77ac30}{(ii)} & \textcolor[HTML]{0072bd}{(iv)}        & \textcolor[HTML]{505050}{(i)}  & $\widehat{a}_{k}$ & $a$ \\
        (C) & \textcolor[HTML]{77ac30}{(ii)} & \textcolor[HTML]{0072bd}{(iv)}        & \textcolor[HTML]{d95319}{(iii)}& $\widehat{a}_{k}$ & $\breve{a}_{k}$ \\
        (D) & \textcolor[HTML]{0072bd}{(iv)} & \textcolor[HTML]{d95319}{(iii)}       & \textcolor[HTML]{77ac30}{(ii)} & $\breve{a}_{k}$ & $\widehat{a}_{k}$ \\
        (E) & \textcolor[HTML]{77ac30}{(ii)} & \textcolor[HTML]{d95319}{(iii)}, \textcolor[HTML]{0072bd}{(iv)}  & \textcolor[HTML]{77ac30}{(ii)} & $\widehat{a}_{k}$ & $\widehat{a}_{k+1}$ \\
        \bottomrule
    \end{tabularx}
\end{table}

\subsection{Modification of Phase Generation}
\label{ssec:phase}

Variants of the PV, see Sec.\:\ref{ssec:PV}, usually employ phase generation methods  tailored to a uniform hop size. However, it is straightforward to adapt formulas Eq.~\eqref{eq:HPD} and Eq.~\eqref{eq:RPPformula}, so long as the number of frequency bins is equal to a constant $M$ for all time frames, as considered here. Simply exchange $a$ for $a_n$ in Eq.~\eqref{eq:HPD} and $\tilde{a}$ for $\lceil \alpha a_n \rceil$ in Eq.~\eqref{eq:RPPformula}. Other standard techniques, such as identity phase locking, remain directly applicable. 

In principle, it is possible to vary the number of frequency bins $M$ at each time step as well, e.g., proportional to the window length to further reduce oversampling. However, doing so would require both the use of directional derivatives and oblique integration paths, introducing significant computational (and book-keeping) overhead, as well as implementation complexity. Furthermore, the quality of PV-based time-stretching benefit from a high number of frequency channels (i.e., high oversampling) in our experience. Overall, the (moderate) reduction in oversampling may be outweighed by the loss in audio quality and increased complexity of the implementation.

\section{Considerations for Bounded-delay, Online Implementation}
\label{sec:on-line}

\begin{figure}[t]
    \centering
    \includegraphics[width=1\linewidth]{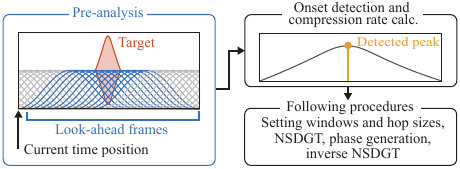}
    \caption{
    Conceptual diagram of a bounded-delay, online implementation. 
    The left panel illustrates the pre-analysis of a target percussive component (red) using the DGT. 
    The left edge represents the current time position, and the analysis windows corresponding to the required look-ahead frames are highlighted in blue. 
    The top-right panel depicts the onset detection and compression rate calculation (corresponding to Fig.\:\ref{fig:compressionRate}\,(f)), displaying the magnitude ratio in gray and the detected peak in yellow. 
    The bottom-right panel outlines the subsequent procedures.
    }
    \label{fig:online}
\end{figure}

Although the present paper and the associated implementation of \mbox{SELEBI} focus on offline processing, we briefly discuss the requirements for a bounded-delay, online implementation. 
As illustrated in Fig.\:\ref{fig:online}, the primary factor determining the feasibility of such an implementation is the look-ahead required by the pre-analysis stage for onset detection and compression rate estimation (Sec.\:\ref{ssec:compressionRate}). 
Accurately capturing these percussive components requires a sufficient observation buffer. 
The left panel of Fig.\:\ref{fig:online} depicts this pre-analysis stage using a standard DGT. 
To provide the necessary context for identifying local maxima as the amplitude evolves (top-right panel), this buffer needs to span approximately twice the maximum analysis window length of the adapted \mbox{NSDGT}, as can be seen in the left panel. 
Once the peak is detected and the compression rate is calculated, the subsequent steps, including adaptive windowing, phase generation, and the inverse \mbox{NSDGT}, can be seamlessly executed on the buffered frames. 
Consequently, the total algorithmic latency is primarily bounded by twice the maximum analysis window length. 
While this estimation assumes a conservative buffer size, future improvements in peak detection, such as exploiting causal temporal features, could potentially reduce this look-ahead requirement, further minimizing the system delay.

\section{Experiments}

\begin{table*}[t]
\caption{Comparison of error measurements for synthetic signals (Sec.\:\ref{ssec:synth}).
The top and bottom rows show the results of $2\times$ and $4\times$ stretching, respectively.
``DIALGA (PVDR)'' denotes DIALGA with PVDR phase generation, and similarly for other methods.
The best and second-best scores are in \textbf{bold} with blue and light blue backgrounds, respectively.}
\centering
\begin{tabularx}{0.89\textwidth}{@{}cl|YYYYY|YY@{}} \toprule
\multirow{2}{*}{} & & PV & PVDR & PVHP & DIALGA & DIALGA & \mbox{SELEBI} & \mbox{SELEBI} \\ 
& &  &  &  & (PV) & (PVDR) & (PV) & (PVDR)\\ \midrule
\multirow{5}{5pt}{\rotatebox[origin=c]{90}{$\alpha=2$}}& Impulse  & 0.2357 & 0.2402 & 0.6993 & \cellcolor{softBlue}\underline{\textbf{0.1009}} & \cellcolor{softSoftBlue}\textbf{0.1110} & 0.2326 & 0.2326 \\
& Sinusoid + Impulse   & 0.0113 & 0.0091 & 0.0130 & \cellcolor{softSoftBlue}\textbf{0.0090} & 0.0092 & 0.0106 & \cellcolor{softBlue}\underline{\textbf{0.0051}}  \\
& Harmonic Sinusoid + Impulse & 0.0129 & 0.0107 & 0.0139 & 0.0136 & \cellcolor{softSoftBlue}\textbf{0.0107} & 0.0128 & \cellcolor{softBlue}\underline{\textbf{0.0098}} \\ 
& Transient & 0.5806 & 0.5666 & 0.5924 & 0.5659 & \cellcolor{softSoftBlue}\textbf{0.5647} & 0.5791 & \cellcolor{softBlue}\underline{\textbf{0.5552}} \\ 
& Sinusoid + Transient & 0.2506 & 0.2446 & 0.4244 & 0.2455 & \cellcolor{softSoftBlue}\textbf{0.2445} & 0.2521 & \cellcolor{softBlue}\underline{\textbf{0.2414}} \\ \midrule\midrule
\multirow{5}{0pt}{\rotatebox[origin=c]{90}{$\alpha=4$}}&Impulse & 0.3080 & 0.3080 & 1.3679 & \cellcolor{softBlue}\underline{\textbf{0.0559}} & \cellcolor{softBlue}\underline{\textbf{0.0559}} & \cellcolor{softSoftBlue}\textbf{0.3069} & \textbf{0.3069} \\
& Sinusoid + Impulse   & 0.0220 & \cellcolor{softBlue}\underline{\textbf{0.0149}} & 0.0242 & 0.0267 & 0.0168 & 0.0206 & \cellcolor{softSoftBlue}\textbf{0.0151}  \\
& Harmonic Sinusoid + Impulse & 0.0298 & \cellcolor{softBlue}\underline{\textbf{0.0264}} & 0.0307 & 0.0315 & 0.0270 & 0.0334 & \cellcolor{softSoftBlue}\textbf{0.0265} \\ 
& Transient & 1.3835 & 1.3153 & 1.3838 & 1.4153 & \cellcolor{softSoftBlue}\textbf{1.3097} & 1.3812 & \cellcolor{softBlue}\underline{\textbf{1.2266}}  \\
& Sinusoid + Transient & 0.4746 & 0.4468 & 0.6175 & 0.4585 & \cellcolor{softSoftBlue}\textbf{0.4407} & 0.4719 & \cellcolor{softBlue}\underline{\textbf{0.4316}} \\
\bottomrule
\end{tabularx}
\label{tab:resultSynth}
\end{table*}

In this section, we evaluate the quality of audio signals processed with \mbox{SELEBI} by comparing it with several reference methods. 
The evaluation uses both synthetic and real-world audio signals. 
For synthetic signals, we employ objective metrics to compare the time-stretched results against a ground truth, whereas for real-world signals, we conduct subjective listening tests.
Audio demos are available in our demo page\footnote{\url{https://natsukiakaishi.github.io/selebi/}}.

The sampling frequency for all signals used in the experiments was 22\,050\,Hz\footnote{Corresponding to the native sampling rate of the dataset of real world recordings considered in Experiment B.}.
The DGT used for the PV implementation utilized a Hann window with a length of $L=2^{11}$ samples (93\,ms) and a synthesis hop size of $2^7$ samples (5.8\,ms). 
The number of frequency channels $M$ was set to $L\alpha$ (i.e., $2^{11}\alpha$), a configuration recommended for better synthesis quality.
In accordance with the DGT settings, the longest window length in the \mbox{NSDGT} of the proposed method was set to $V=2^{11}$. 
 The masking threshold parameters of the proposed method were set to $\theta_{\text{mag}}=0.01$, $\underline{\theta_{\text{p}}}=0.5$, and $\overline{\theta_{\text{p}}}=0.75$, based on informal pre-tests.

\subsection{Experiment A: Synthetic Signals}
\label{ssec:synth}

We use idealized, synthetic signals comprised of distinct characteristic combinations of pulses and sinusoids, to study fundamental patterns in and differences between the proposed and reference methods. 
These signals were designed such that an ideal ``ground truth'' is available for objective evaluation. 
In every case and configuration, we applied time-stretching and evaluated the synthesis accuracy based on the relative RMS difference of magnitude spectrograms, i.e., in the T-F domain.

\subsubsection{Test Signals}

We performed time-stretching on five characteristic synthetic signals using relative amplitudes defined with respect to the transient peak.
The five signals were: (1) a unit impulse; (2) a mixture of an impulse and a $1000$\,Hz sinusoid scaled by a factor of $0.5$; (3) a mixture of an impulse and harmonic sinusoids at $1000$, $2000$, and $3000$\,Hz with amplitudes scaled by $0.5$, $0.25$, and $0.125$, respectively; (4) an exponentially decaying transient (based on a $50$\,Hz sinusoid); and (5) the same transient combined with a $1000$\,Hz sinusoid with half the peak amplitude of the transient.
To define the ground truth for the transient preservation quality, the ideal time-stretched signals were constructed by placing the transient components at the new time positions while maintaining their original durations and decay envelopes (i.e., without time-stretching the transients themselves).

\subsubsection{Experimental Conditions}
For this experiment, we considered time-stretching by factors of 2 and 4. For each condition, we compared the following methods: classical PV with identity phase locking%
\footnote{For impulse-only samples, the absence of sinusoidal components means that applying identity phase locking is equivalent to not applying it.} \cite{laroche2002improved} as a baseline, PVDR \cite{doneRight} as a percussion-agnostic state-of-the-art method, HPSS-based time stretching (PVHP) \cite{driedger2013improving} as a standard percussive-aware method, our previous method (DIALGA) \cite{akaishi2023}, and the proposed method (\mbox{SELEBI}). 
For both DIALGA and \mbox{SELEBI}, we evaluated versions using both classical PV and PVDR for phase generation.
All methods, with the exception of PVHP, were implemented in MATLAB using the LTFAT toolbox \cite{ltfat}; for PVHP, we used the authors' original code \cite{driedger2014tsm}.

\subsubsection{Evaluation}

For the evaluation, we use the following spectral error, calculated from the DGT coefficients of the ground truth signal ($\mathbf{X}^{\text{perf}}$) and the stretched signal ($\mathbf{X}$):
\begin{equation}\label{eq:objerror}
    E(\mathbf{X}^{\text{perf}},\mathbf{X}) = \frac{\||\mathbf{X}^{\text{perf}}|-|\mathbf{X}|\|_{\text{F}}}{\||\mathbf{X}^{\text{perf}}|\|_{\text{F}}}.
\end{equation}
The DGT coefficients for evaluation utilized the same window settings ($W$ and $\widetilde{a}$) as the synthesis process. To mitigate boundary artifacts (due to signal truncation and circularity), we evaluated only the time interval corresponding to the percussive components.

\subsubsection{Results}

\begin{figure*}
    \centering
    \includegraphics[width=1\linewidth]{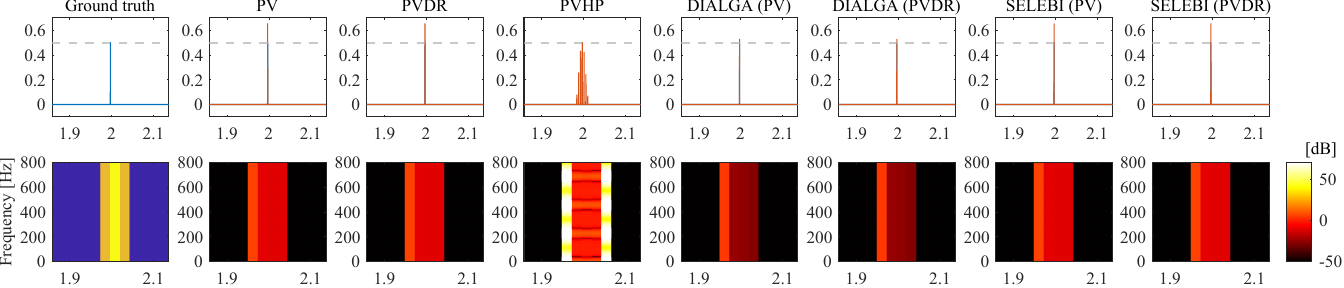}
    \caption{
    An impulse stretched $4\times$. 
    The leftmost column shows the ground truth (blue), while the remaining columns display the stretched signal for each comparison method (red). 
    The figure is divided into two rows: the top row shows a magnified view of the impulse, and the bottom row displays the relative error spectrogram (log scale). 
    In the top row, for visual clarity, the ground truth is plotted as a gray dashed line. In the error spectrogram (bottom), brighter colors indicate larger errors.
    }
    \label{fig:resultImpulse}
\end{figure*}

\begin{figure*}
    \centering
    \includegraphics[width=1\linewidth]{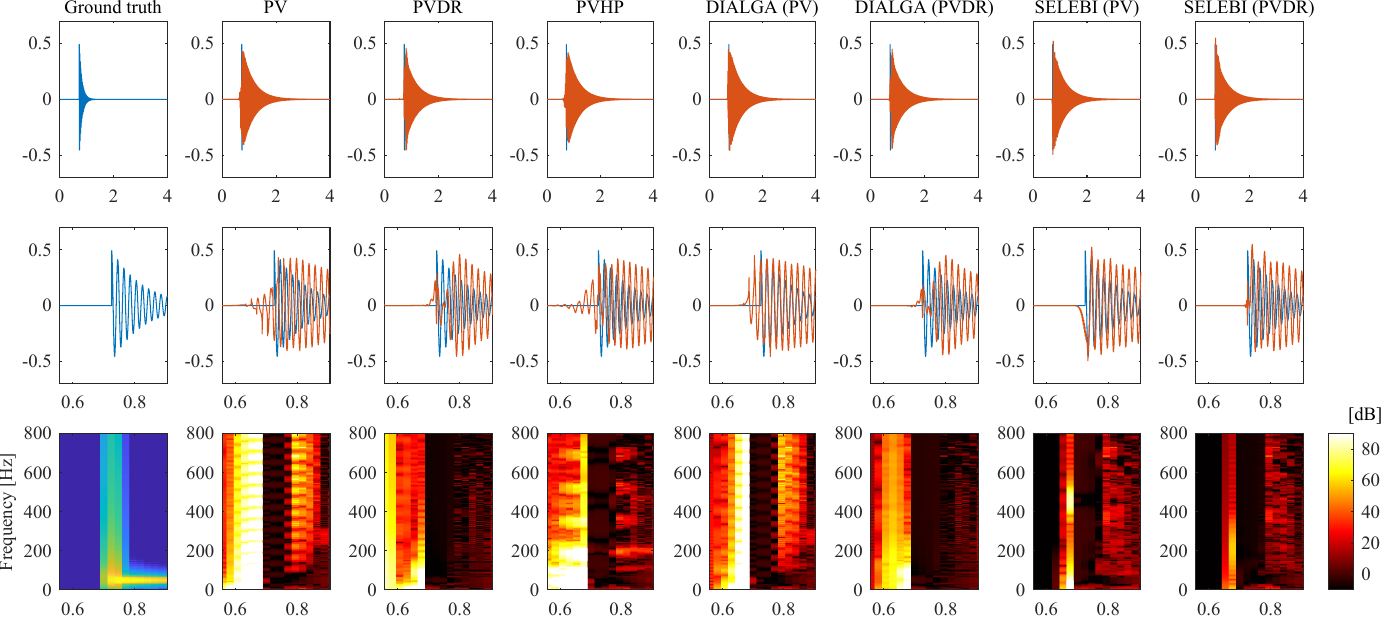}
    \caption{
    Example of a transient signal stretched $4\times$. 
    The leftmost column shows the ground truth (blue), while the remaining columns display the stretched signal for each comparison method (red). 
    The figure is divided into three rows: the top row shows the entire time waveform, the middle row shows a magnified view of the waveform onset, and the bottom row displays the relative error spectrogram (log scale). 
    The horizontal axis for all panels is time (seconds). 
    In the error spectrogram (bottom), brighter color indicates greater error.
    }
    \label{fig:resultSynth}
\end{figure*}

Table~\ref{tab:resultSynth} summarizes the results of experiment. 
\mbox{SELEBI} (PVDR) consistently achieved the lowest or second-lowest error for almost all signals in both $2\times$ and $4\times$ stretching scenarios.
For the simple impulse signal (as shown in Fig.\;\ref{fig:resultImpulse}), even standard PV and PVDR maintained relatively low errors. In contrast, PVHP suffered from performance degradation; this was caused by the WSOLA algorithm (used for stretching of percussive components), which introduced impulse duplication artifacts. 
Notably, DIALGA performed slightly better than \mbox{SELEBI}, the proposed method, on this signal. 
As shown in Fig.\:\ref{fig:resultImpulse}, DIALGA yielded more accurate amplitude values relative to the ground truth.
We attribute the mild degradation in \mbox{SELEBI} to its adaptive window amplitude modulation, which introduces a minor amplification of the impulse amplitude, thereby leading to larger values of the error measure Eq.~\eqref{eq:objerror}.
On the other hand, for more complex signals, 
\mbox{SELEBI} outperformed, or at least nearly matched, the best reference method.

In Fig.\:\ref{fig:resultSynth}, we show the results for the $4\times$-stretched transient signal, where the distinct characteristics of all methods are most apparent.
Due to its sinusoidal characteristic, the decaying tail of this signal is not treated as percussive by any of the considered methods. Hence, the extended decay time visible in the figure is expected, and we observe similar behavior in all cases. However, significant differences can be observed in the attack portion of the stretched signals. 
For classical PV, significant leakage is evident in both the waveform (middle row) and the error spectrogram (bottom row).
PVDR mitigates this leakage to some extent, but residual artifacts remain. In particular, the attack portion of the signal is notably suppressed or delayed. 
PVHP, on the other hand, exhibits leakage stemming from its underlying PV processing; additionally, small amplitude cancellations are observed, likely caused by interference when the processed attack component is superimposed back onto the waveform.
Both variants of DIALGA, based on standard PV and PVDR, respectively, visibly reduce leakage compared to the respective standalone baselines.
\mbox{SELEBI}, in contrast, suppresses leakage almost entirely and preserves the sharp attack of the original signal, providing the best overall result. 

\subsection{Experiment B: Real World Signals}

Although the synthetic signals covered in the previous section facilitate a study of the performance and properties of \mbox{SELEBI} in an idealized setting, Experiment A does not allow us to draw conclusions for its performance on real audio recordings. 
Hence, we evaluated the perceptual quality of \mbox{SELEBI} on a set of real-world recordings, which exhibit greater complexity than the synthetic examples.

\subsubsection{Test Signals} 
We performed time-stretching on seven excerpts from the TSM toolbox \cite{driedger2014tsm}. 
Specifically, we extracted a one-second clip each from the recordings labeled as ``Bongo', ``CastanetsViolin'', ``DrumSolo'', ``Glockenspiel'', ``Jazz'', ``Pop'', and ``Stepdad''. 
These excerpts capture characteristic features of the recordings, e.g., for the excerpt from ``Jazz'' sample, we chose a section with a balanced mix of brass, piano, kick drum, and snare drum. Fig.\:\ref{fig:examReal} shows two representative examples from the set of test signals and the resulting signals after stretching by a factor of $4$: ``CastanetsViolin'' and ``Glockenspiel,'' with specific regions of interest enlarged for clarity. 

\subsubsection{Experimental Conditions}

The tested conditions largely follow the previous experiment, considering time-stretching by factors of 2 and 4. 
We consider the same methods as before, except for the omission of the DIALGA and \mbox{SELEBI} variants based on the classical PV, to reduce the number of conditions for the subjective test. 
In preliminary tests, these methods never yielded a clear advantage over their PVDR-based counterparts, justifying the omission.

\subsubsection{Evaluation}

\begin{figure*}
    \centering
    \includegraphics[width=1\linewidth]{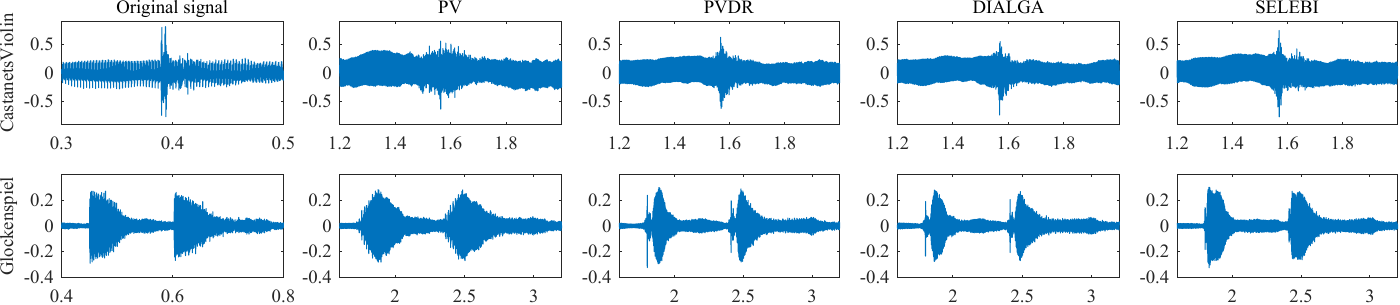}
    \caption{
    Examples of the stretched signals of ``CastanetsViolin'' and ``Glockenspiel''.
    The leftmost column shows the original signal, while the remaining columns display the stretched signal for each comparison method.
    The horizontal axis for all panels is time (seconds). Transient smearing in classical PV is apparent in both examples, whereas PVDR and DIALGA demonstrate different, characteristic artifacts at percussive events in the synthesized signals: PVDR exhibits slight smearing and both methods result in somewhat reduced amplitude of attacks. Additionally, in the ``Glockenspiel'' excerpt PVDR and DIALGA show a separation of the transient attack portion from the more harmonic decay segment of each played note. In contrast, \mbox{SELEBI} yields a more natural waveform that preserves attacks and harmonic signal components and more closely resembles the original signal. 
    }
    \label{fig:examReal}
\end{figure*}

\begin{figure}
    \centering
    \includegraphics[width=1\linewidth]{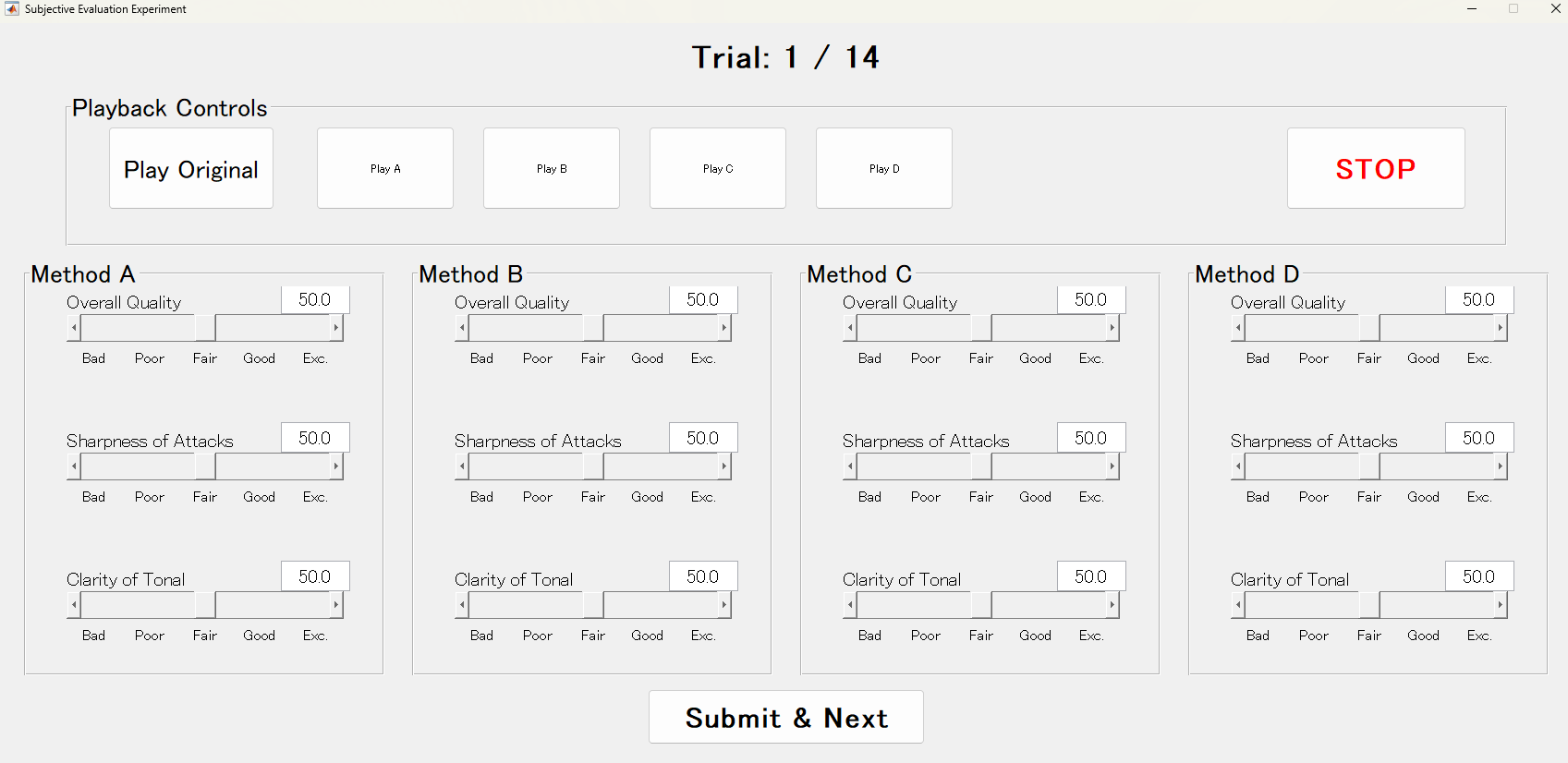}
    \caption{
    Screenshot of the UI of the code used for subjective evaluation.
    }
    \label{fig:UI}
\end{figure}

We conducted a subjective evaluation based on the ITU-R BS.2132-0 recommendation \cite{eval}, a method suitable for scenarios where no reference (ground truth) signals are available%
\footnote{This recommendation allows for evaluation even when a standard anchor is unavailable. In this experiment, no explicit anchor was used, although the standard PV could technically serve as one.}, as is the case for real-world time-stretching.
Ten subjects with normal hearing listeners (i.e., having no known hearing impairment) participated in the experiment.
At the beginning of the session, subjects were instructed to adjust the playback volume to a comfortable listening level. 
They were informed that the test sounds included the original signal as well as its $2\times$ and $4\times$ time-stretched versions.
The four processing methods were presented blindly in a randomized order.
Using a MATLAB-based interface (Fig.\:\ref{fig:UI}), participants were instructed to rate the sounds relative to the original signal based on three criteria: ``Overall Quality,'' ``Sharpness of Attacks,'' and ``Clarity of Tonal Components.'' 
The ratings were recorded on a continuous scale from 0 to 100, divided into five quality intervals: ``Bad'' (0--20), ``Poor'' (20--40), ``Fair'' (40--60), ``Good'' (60--80), and ``Excellent'' (80--100).
They were permitted to listen to the samples as many times as they wished.
The experiment consisted of 14 trials in total: 7 excerpts with $2\times$ stretching followed by 7 excerpts with $4\times$ stretching.

\subsubsection{Results}

\begin{table*}[t]
\caption{Comparison of the mean scores across three metrics: Overall Quality, Sharpness of Attacks, and Clarity of Tonal Components.
The best and second-best scores are in \textbf{bold} with blue and light blue backgrounds, respectively.}
\centering
\fontsize{7.5}{9}\selectfont 
\begin{tabularx}{0.98\textwidth}{@{}cl|YYYY|YYYY|YYYY@{}}
\toprule
& & \multicolumn{4}{c|}{\textbf{Overall Quality}} & \multicolumn{4}{c|}{\textbf{Sharpness of Attacks}} & \multicolumn{4}{c}{\textbf{Clarity of Tonal Components}} \\
\cmidrule(lr){3-6} \cmidrule(lr){7-10} \cmidrule(lr){11-14}
& & PV & PVDR & DIALGA & \mbox{SELEBI} & PV & PVDR & DIALGA & \mbox{SELEBI} & PV & PVDR & DIALGA & \mbox{SELEBI} \\
\midrule
\multirow{8}{5pt}{\rotatebox[origin=c]{90}{$\alpha=2$}}
& Bongo & 14.98 & \cellcolor{softSoftBlue}\textbf{82.48} & 73.28 & \cellcolor{softBlue}\underline{\textbf{94.18}} & 19.62 & \cellcolor{softSoftBlue}\textbf{85.92} & 72.88 & \cellcolor{softBlue}\underline{\textbf{92.31}} & 15.71 & \cellcolor{softSoftBlue}\textbf{77.22} & 69.11 & \cellcolor{softBlue}\underline{\textbf{85.72}} \\
& CastanetsViolin & 28.77 & 72.73 & \cellcolor{softBlue}\underline{\textbf{80.69}} & \cellcolor{softSoftBlue}\textbf{78.78} & 10.01 & 69.17 & \cellcolor{softSoftBlue}\textbf{71.35} & \cellcolor{softBlue}\underline{\textbf{75.38}} & 50.33 & 72.27 & \cellcolor{softBlue}\underline{\textbf{81.55}} & \cellcolor{softSoftBlue}\textbf{78.32} \\
& DrumSolo & 17.43 & 56.48 & \cellcolor{softSoftBlue}\textbf{58.07} & \cellcolor{softBlue}\underline{\textbf{64.49}} & 10.56 & \cellcolor{softSoftBlue}\textbf{54.93} & 54.49 & \cellcolor{softBlue}\underline{\textbf{59.88}} & 22.74 & \cellcolor{softSoftBlue}\textbf{57.35} & 51.17 & \cellcolor{softBlue}\underline{\textbf{57.69}} \\
& Glockenspiel & 33.57 & \cellcolor{softSoftBlue}\textbf{67.24} & 63.93 & \cellcolor{softBlue}\underline{\textbf{74.21}} & 29.48 & 63.26 & \cellcolor{softSoftBlue}\textbf{63.73} & \cellcolor{softBlue}\underline{\textbf{73.26}} & 40.28 & \cellcolor{softSoftBlue}\textbf{67.59} & 66.47 & \cellcolor{softBlue}\underline{\textbf{74.66}} \\
& Jazz & 33.57 & \cellcolor{softSoftBlue}\textbf{68.26} & 63.71 & \cellcolor{softBlue}\underline{\textbf{68.58}} & 28.56 & \cellcolor{softSoftBlue}\textbf{62.05} & 61.80 & \cellcolor{softBlue}\underline{\textbf{68.65}} & 33.92 & \cellcolor{softBlue}\underline{\textbf{67.63}} & \cellcolor{softSoftBlue}\textbf{64.59} & 62.09 \\
& Pop & 29.93 & \cellcolor{softBlue}\underline{\textbf{78.76}} & \cellcolor{softSoftBlue}\textbf{76.29} & 74.61 & 21.30 & \cellcolor{softSoftBlue}\textbf{73.34} & \cellcolor{softBlue}\underline{\textbf{74.17}} & 69.54 & 35.32 & \cellcolor{softBlue}\underline{\textbf{77.95}} & 74.62 & \cellcolor{softSoftBlue}\textbf{76.45} \\
& Stepdad & 30.78 & \cellcolor{softBlue}\underline{\textbf{72.09}} & \cellcolor{softSoftBlue}\textbf{68.63} & 68.25 & 26.71 & \cellcolor{softBlue}\underline{\textbf{71.79}} & \cellcolor{softSoftBlue}\textbf{69.92} & 63.98 & 30.71 & \cellcolor{softSoftBlue}\textbf{67.04} & \cellcolor{softBlue}\underline{\textbf{67.83}} & 61.62 \\
\cmidrule(lr){2-14}
& Average & 27.00 & \cellcolor{softSoftBlue}\textbf{71.15} & 69.23 & \cellcolor{softBlue}\underline{\textbf{74.73}} & 20.89 & \cellcolor{softSoftBlue}\textbf{68.64} & 66.91 & \cellcolor{softBlue}\underline{\textbf{71.86}} & 32.71 & \cellcolor{softSoftBlue}\textbf{69.58} & 67.91 & \cellcolor{softBlue}\underline{\textbf{70.93}} \\
\midrule\midrule
\multirow{8}{5pt}{\rotatebox[origin=c]{90}{$\alpha=4$}}
& Bongo & 7.19 & 67.83 & \cellcolor{softSoftBlue}\textbf{69.07} & \cellcolor{softBlue}\underline{\textbf{87.59}} & 5.89 & 73.46 & \cellcolor{softSoftBlue}\textbf{74.68} & \cellcolor{softBlue}\underline{\textbf{88.47}} & 13.14 & 61.25 & \cellcolor{softSoftBlue}\textbf{67.37} & \cellcolor{softBlue}\underline{\textbf{80.75}} \\
& CastanetsViolin & 25.23 & \cellcolor{softSoftBlue}\textbf{71.86} & 67.95 & \cellcolor{softBlue}\underline{\textbf{74.03}} & 9.56 & \cellcolor{softSoftBlue}\textbf{69.66} & 66.41 & \cellcolor{softBlue}\underline{\textbf{73.51}} & 41.17 & \cellcolor{softBlue}\underline{\textbf{78.10}} & 72.79 & \cellcolor{softSoftBlue}\textbf{74.43} \\
& DrumSolo & 5.29 & 42.33 & \cellcolor{softSoftBlue}\textbf{44.11} & \cellcolor{softBlue}\underline{\textbf{57.56}} & 3.61 & \cellcolor{softSoftBlue}\textbf{40.08} & 39.60 & \cellcolor{softBlue}\underline{\textbf{58.01}} & 11.75 & 44.03 & \cellcolor{softSoftBlue}\textbf{48.53} & \cellcolor{softBlue}\underline{\textbf{53.67}} \\
& Glockenspiel & 26.96 & \cellcolor{softSoftBlue}\textbf{70.55} & 63.28 & \cellcolor{softBlue}\underline{\textbf{79.05}} & 20.24 & \cellcolor{softSoftBlue}\textbf{65.85} & 60.62 & \cellcolor{softBlue}\underline{\textbf{79.68}} & 36.01 & \cellcolor{softSoftBlue}\textbf{67.56} & 66.00 & \cellcolor{softBlue}\underline{\textbf{77.46}} \\
& Jazz & 23.67 & \cellcolor{softSoftBlue}\textbf{58.26} & 56.04 & \cellcolor{softBlue}\underline{\textbf{78.64}} & 16.57 & \cellcolor{softSoftBlue}\textbf{58.55} & 52.70 & \cellcolor{softBlue}\underline{\textbf{79.35}} & 27.98 & 57.49 & \cellcolor{softSoftBlue}\textbf{57.60} & \cellcolor{softBlue}\underline{\textbf{73.16}} \\
& Pop & 21.01 & 57.78 & \cellcolor{softSoftBlue}\textbf{60.79} & \cellcolor{softBlue}\underline{\textbf{76.13}} & 15.34 & 54.14 & \cellcolor{softSoftBlue}\textbf{58.68} & \cellcolor{softBlue}\underline{\textbf{79.50}} & 23.15 & 57.82 & \cellcolor{softSoftBlue}\textbf{62.00} & \cellcolor{softBlue}\underline{\textbf{73.99}} \\
& Stepdad & 17.44 & \cellcolor{softBlue}\underline{\textbf{66.55}} & \cellcolor{softSoftBlue}\textbf{64.32} & 63.97 & 9.72 & \cellcolor{softSoftBlue}\textbf{63.11} & \cellcolor{softBlue}\underline{\textbf{63.76}} & 61.62 & 26.23 & \cellcolor{softBlue}\underline{\textbf{68.00}} & 62.07 & \cellcolor{softSoftBlue}\textbf{62.23} \\
\cmidrule(lr){2-14}
& Average & 18.11 & \cellcolor{softSoftBlue}\textbf{62.17} & 60.79 & \cellcolor{softBlue}\underline{\textbf{73.85}} & 11.56 & \cellcolor{softSoftBlue}\textbf{60.69} & 59.49 & \cellcolor{softBlue}\underline{\textbf{74.31}} & 25.63 & 62.04 & \cellcolor{softSoftBlue}\textbf{62.34} & \cellcolor{softBlue}\underline{\textbf{70.81}} \\
\bottomrule
\end{tabularx}
\label{tab:resultReal}
\end{table*}

Table~\ref{tab:resultReal} presents the results of the subjective evaluation. 
Regarding ``Overall Quality,'' \mbox{SELEBI} achieved the highest scores in most conditions. 
Focusing on the mean scores, \mbox{SELEBI} outperformed both PVDR and DIALGA by approximately 3 points for the $2\times$ stretch and 10 points for the $4\times$ stretch. 
A similar trend was observed for ``Sharpness of Attacks,'' where \mbox{SELEBI} led by approximately 2 points ($2\times$) and 14 points ($4\times$). 
Regarding ``Clarity of Tonal Components,'' \mbox{SELEBI} was comparable to the references at $2\times$ but scored about 8 points higher at $4\times$. 
These results, which confirm the visual observation (Fig.\;\ref{fig:examReal}), indicate that \mbox{SELEBI} effectively preserves percussive components without compromising, and in fact improving, the perception of tonal components. This advantage was more pronounced at higher stretch factors.

Examining individual samples reveals further insights. 
\mbox{SELEBI} demonstrated a particularly large margin of improvement on simpler excerpts such as ``Bongo,'' ``Glockenspiel,'' and ``Drum Solo.'' 
In terms of ``Overall Quality'' and ``Sharpness of Attacks,'' DIALGA outperforms PVDR on simpler signals (e.g., ``CastanetsViolin'' at $2\times$ stretch and ``Bongo'' at $4\times$ stretch), but is at a disadvantage on more complex excerpts (e.g., ``Jazz,'' ``Pop,'' and ``Stepdad''), resulting in similar average ratings overall. 
Consequently, the overall average ratings are comparable.
\mbox{SELEBI} achieves  excellent results for all conditions, demonstrating its robustness in preserving percussive components within complex mixtures, although on the most complex signals ``Pop,'' and ``Stepdad'', we observed slightly decreased performace over PVDR and DIALGA for some cases and criteria.

\section{Conclusion}

In this paper, we proposed \mbox{SELEBI} to mitigate percussion smearing in phase vocoder-based time stretching.
To resolve the fundamental magnitude-phase mismatch, we introduced a ``magnitude squeezing'' approach that leverages the \mbox{NSDGT} to directly obtain temporally compressed magnitude representations, offering a rigorous mathematical improvement over the previous heuristic method, DIALGA.
Our method achieves this by adaptively controlling both window lengths and hop sizes.
The adaptive windowing relies not only on onset decisions but is also guided by a compression rate based on the energy ratio of percussive components, which ensures that the preserved percussive components sound perceptually natural.
Guided by this rate, the analysis window lengths are reduced around percussive events and linearly transitioned to standard sizes to prevent abrupt artifacts.
Furthermore, to reduce computational complexity and guarantee numerical stability within these short-window regions, we adaptively change the hop size.
By smoothly interpolating between short hop sizes for transients and longer hop sizes for other components, \mbox{SELEBI} realizes a mathematically stable \mbox{NSDGT}.

Both objective and subjective evaluations confirmed the practical efficacy of this framework.
\mbox{SELEBI} significantly outperformed conventional methods (e.g., PVDR and DIALGA), particularly at an extreme $4\times$ stretch.
It successfully restored transient attack sharpness while preserving, and often even improving, tonal clarity.
This advantage was most pronounced in simpler excerpts. 
While the method proved robust across various conditions, evaluations on highly dense polyphonic mixtures indicated that performance gains can occasionally plateau, performing comparably to reference methods.
This highlights the inherent difficulty of adaptive windowing in saturated time-frequency representations.
Future work will focus on addressing these edge cases, as well as transitioning the current offline architecture into a bounded-delay, online implementation for real-time applications.

\section*{Acknowledgment}

The work of N.~Holighaus was supported by the Austrian Science Fund (FWF) project DISCO [10.55776/PAT4780023]. N.~Holighaus would like to thank K.~Yatabe and Tokyo University of Agriculture and Technology (TUAT) for their hospitality during two extended visits in 2025, both kindly funded by TUAT.

\bibliographystyle{IEEEtran}
\bibliography{Bibliography}

\end{document}